\documentclass[11pt]{article}

\input epsf
\usepackage{latexsym}
\usepackage{amssymb}
\usepackage{amsmath}
\usepackage[dvips]{graphicx}
\usepackage{array}

\def\Bc{\vec{\cal B}_k}
\def\Ec{\vec{\cal E}_k}

\begin{document}
\centerline{\Large \bf  Primordial magnetic fields and nonlinear 
electrodynamics}

\vskip 2 cm

\centerline{Kerstin E. Kunze
\footnote{E-mail: kkunze@usal.es, Kerstin.Kunze@cern.ch} }

\vskip 0.3cm

\centerline{{\sl Departamento de F\'\i sica Fundamental,}}
\centerline{and}
\centerline{{\sl Instituto Universitario de F\'\i sica Fundamental y 
Matem\'aticas (IUFFyM),}}
\centerline{{\sl Universidad de Salamanca,}}
\centerline{{\sl Plaza de la Merced s/n, E-37008 Salamanca, Spain }}

\vskip 1.5cm

\centerline{\bf Abstract}
\vskip 0.5cm
\noindent
The creation of large scale magnetic fields is studied 
in an inflationary universe where electrodynamics is 
assumed to be nonlinear. 
After inflation ends electrodynamics becomes linear
and thus the description of reheating and the subsequent
radiation dominated stage are unaltered.
The nonlinear regime of electrodynamics 
is described by lagrangians
having a power law dependence on one of the invariants of the 
electromagnetic field.
It is found that there is a range of parameters for which primordial 
magnetic fields of cosmologically interesting strengths can 
 be created.

\vskip 1cm

\section{Introduction}

Magnetic fields are observed to be associated with 
most structures in the universe. Observations indicate 
magnetic fields on stellar upto supergalactic scales.
The field strengths vary from a few $\mu$G on galactic
scale, upto $10^3$ G for solar type stars and upto 
$10^{13}$ G for neutron stars.
Furthermore, the magnetic field
structure depends on the object it is associated with. Thus, e.g.,
magnetic fields observed in elliptical galaxies show a 
different structure from those associated with 
spiral galaxies \cite{mag}. 

Magnetic fields in stars can be explained by 
the formation of protostars out of condensed
interstellar matter which was pervaded by a  
pre-existing large scale magnetic field (see, e.g.,
\cite{rees}). 
An open problem remains to explain the origin of 
such large scale magnetic fields.

There are different types of proposals.
Ranging from processes on small scales, such as 
vortical perturbations and phase transitions
to models taking advantage of the possibility of 
amplifying perturbations in the electromagnetic 
field during inflation in the early universe
(see, e.g., \cite{rev-mag}).

Inflation provides a mechanism to amplify perturbations 
in some field to appreciable size. In order for this mechanism to 
lead to primordial magnetic seed fields of cosmologically 
interesting strength, the corresponding lagrangian 
should not be conformally invariant.
The Maxwell lagrangian describing linear electrodynamics is 
conformally invariant. There have been already a multitude of 
proposals to break the conformal invariance of the Maxwell 
theory \cite{tw}, e.g. by coupling to a scalar field \cite{mag-sc}, 
breaking Lorentz invariance \cite{mag-lor},
adding extra dimensions \cite{mag-ex} or a coupling to curvature terms
\cite{mag-curv}.

Here nonlinear electrodynamics is considered. 
It has its origins in the search for a classical singularity-free
theory of the electron by Born and Infeld \cite{bi}. Later on 
it was realized that virtual electron pair creation 
induces a self-coupling of the electromagnetic field.
For slowly varying, but arbitrarily strong electromagnetic fields
the self-interaction energy was computed by Heisenberg and Euler
(cf. \cite{he}-\cite{bb}).  

The propagation of a photon in an external electromagnetic field
can be described effectively by the Heisenberg-Euler 
langrangian. Moreover, the transition amplitude 
for photon splitting in quantum electrodynamics is 
nonvanishing in this case. In principle, this might lead to 
observational effects, e.g., on the electromagnetic 
radiation coming from neutron stars 
which are known to have strong magnetic fields.
\cite{bb,p-sp}.
In particular, certain features in the spectra of pulsars 
can be explained by photon splitting \cite{puls}.

Finally, Born-Infeld type actions
also appear as a low energy effective action of open strings 
\cite{bi-strings,gh}.
As was shown in \cite{dbi} the low energy dynamics of D-branes
is described by the Dirac-Born-Infeld action.

The model of the cosmological background that will be considered 
consists of a stage of de Sitter inflation followed by reheating and a standard
radiation dominated stage. 
Quantum fluctuations in the electromagnetic field are excited within the
horizon during inflation. Once outside the horizon they become classical 
perturbations.
As mentioned above, in general, the conformal invariance of the four 
dimensional Maxwell field has to be broken in order to amplify
the perturbations in the electromagnetic field significantly.
Thus, here electrodynamics is considered to be 
nonlinear during the de Sitter stage. This could be motivated by the 
presence of possible quantum corrections to quantum electrodynamics 
at high energies. However, once inflation ends
electrodynamics is described by standard Maxwell electrodynamics.
Thus the subsequent evolution described by the standard model of cosmology 
is unchanged.

\section{Nonlinear electrodynamics in the early universe}
\setcounter{equation}{0}

The Born-Infeld or Heisenberg-Euler lagrangians provide  
particular examples of theories of nonlinear electrodynamics.
In general the action of nonlinear electrodynamics coupled minimally
to gravity can be written as, see e.g. \cite{gh,p}
\begin{eqnarray}
S=\frac{1}{16\pi G_N}\int d^4x\sqrt{-g}R+\frac{1}{4\pi}
\int d^4x \sqrt{-g} L(X,Y),
\end{eqnarray}
where $L(X,Y)$ is the lagrangian of nonlinear electrodynamics. Furthermore,
the invariants are denoted by
$X\equiv\frac{1}{4}F_{\mu\nu}F^{\mu\nu}$ and 
$Y\equiv\frac{1}{4}F_{\mu\nu}\;^{*}F^{\mu\nu}$,
where $^{*}F^{\mu\nu}$ is the dual bi-vector given by
\newline
$^{*}F^{\mu\nu}=\frac{1}{2\sqrt{-g}}\epsilon^{\mu\nu\alpha\beta}F_{\alpha\beta}$,
and $\epsilon^{\mu\nu\alpha\beta}$ the Levi-Civita tensor with 
$\epsilon_{0123}=+1$.

The equations of motion are given by
\begin{eqnarray}
\nabla_{\mu}P^{\mu\nu}=0
\label{p1}
\end{eqnarray}
where $P_{\mu\nu}=-\left(L_X F_{\mu\nu}+L_Y \;^{*}F_{\mu\nu}\right)$,
furthermore $L_A$ denotes $L_A=\partial L/\partial A$,
and 
\begin{eqnarray}
\nabla_{\mu} \, ^{*}F^{\mu\nu}=0,
\label{p2}
\end{eqnarray}
which implies that $F_{\mu\nu}=\partial_{\mu}A_{\nu}-\partial_{\nu}A_{\mu}$.
The notation used is given by $\mu,\nu..=0...3$ and $i,j..=1,2,3$. 
Moreover, the electromagnetic field is treated as a perturbation 
so that the vacuum Einstein equations apply to the background cosmology.
The background metric is chosen to be of the form
\begin{eqnarray}
ds^2=a^2(\eta)\left[-d\eta^2+d{\mathbf x}^2\right].
\end{eqnarray}
Furthermore, following \cite{tw} the Maxwell tensor is written 
in terms of the electric and magnetic fields, $\vec{E}$ and $\vec{B}$,
respectively, as follows,
\begin{eqnarray}
F_{\mu\nu}=a^2\left(
\begin{array}{cccc}
0&-E_x&-E_y&-E_z\\
E_x&0&B_z&-B_y\\
E_y&-B_z&0&B_x\\
E_z&B_y&-B_x&0
\end{array}
\right).
\end{eqnarray}
These are the components in the "lab" frame in which in linear 
electrodynamics the frozen-in magnetic field decays in an expanding
universe as $1/a^2$.

Then equations (\ref{p1}) and (\ref{p2}) imply 
\begin{eqnarray}
\nabla\cdot\vec{E}+\frac{(\nabla L_X)\cdot\vec{E}}{L_X}
-\frac{(\nabla L_Y)\cdot\vec{B}}{L_X}=0
\label{p3}\\
\frac{1}{a^2}\partial_{\eta}(a^2\vec{E})-\nabla\times\vec{B}+\frac{\partial_{\eta}L_X}{L_X}
\vec{E}-\frac{\partial_{\eta}L_Y}{L_X}\vec{B}-\frac{(\nabla L_X)\times\vec{B}}{L_X}
-\frac{(\nabla L_Y)\times\vec{E}}{L_X}=0
\label{p4}\\
\nabla\cdot\vec{B}=0
\label{p5}\\
\frac{1}{a^2}\partial_{\eta}(a^2\vec{B})+\nabla\times\vec{E}=0
\label{p6}
\end{eqnarray}
Although  Maxwell's equations are recovered for lagrangians of the form $L=n_0X+n_1Y$,
where $n_0$ and $n_1$ are constants,
standard linear electrodynamics corresponds to $L=-X$.
Equations (\ref{p3}) to (\ref{p6}) is a set of four first order partial differential equations 
which can be transformed into a set of two second order partial differential equations.
This describes the evolution  of the nonlinear electromagnetic field in 
a curved background.  In linear electrodynamics this procedure leads to 
two decoupled wave equations, one for the magnetic and one for the
electric field. In the case of nonlinear electrodynamics the resulting wave equations for the
electric and the magnetic fields are no longer decoupled because of the 
nonlinearities. 

Taking the curl of equation (\ref{p4}) and using equations (\ref{p5}) and (\ref{p6})
a wave type equation for the magnetic field  $\vec{B}$ can be found.
\begin{eqnarray}
& &\frac{1}{a^2}\frac{\partial^2}{\partial\eta^2}(a^2\vec{B})
+\frac{1}{a^2}\frac{\partial_{\eta} L_X}{L_X}\partial_{\eta}(a^2\vec{B})
+\frac{1}{a^2}\frac{\partial_{\eta}L_Y}{L_X}\partial_{\eta}(a^2\vec{E})
+\frac{\partial_{\eta}L_Y}{L_X}\left(\frac{\partial_{\eta}L_X}{L_X}\vec{E}-\frac{\partial_{\eta}L_Y}{L_X}
\vec{B}\right)
\nonumber\\
&-&\Delta\vec{B}
+\vec{E}\times\nabla
\left(\frac{\partial_{\eta}L_X}{L_X}\right)
-\vec{B}\times\nabla\left(\frac{\partial_{\eta}L_Y}{L_X}\right)
-\frac{\partial_{\eta}L_Y}{L_X}
\left[\frac{(\nabla L_X)\times\vec{B}}{L_X}+\frac{(\nabla L_Y)\times\vec{E}}{L_X}\right]
\nonumber\\
&+&\nabla\times\left[\frac{(\nabla L_X)\times\vec{B}}{L_X}\right]
+\nabla\times\left[\frac{(\nabla L_Y)\times\vec{E}}{L_X}\right]=0
\label{b1}
\end{eqnarray}
Similarly, taking the time derivative of of equation (\ref{p4}) and using the 
remaining equations results in 
a wave type equation for the electric field $\vec{E}$,
\begin{eqnarray}
& &\frac{\partial^2}{\partial\eta^2}\left(a^2\vec{E}\right)
+\partial_{\eta}\left[\frac{\partial_{\eta} L_X}{L_X}a^2\vec{E}\right]
-\partial_{\eta}\left[\frac{\partial_{\eta}L_Y}{L_X}a^2\vec{B}\right]
\nonumber\\
&-&
\Delta\left(a^2\vec{E}\right)
-\partial_{\eta}\left[\frac{\left(\nabla L_X\right)\times a^2\vec{B}}{L_X}\right]
-\partial_{\eta}\left[\frac{\left(\nabla L_Y\right)\times a^2\vec{E}}{L_X}\right]
\nonumber\\
&-&\nabla\left[\frac{(\nabla L_X)\cdot(a^2\vec{E})}{L_X}\right]
+\nabla\left[\frac{(\nabla L_Y)\cdot (a^2\vec{B})}{L_X}\right]
=0
\label{ef1}
\end{eqnarray}

Equations (\ref{b1}) and (\ref{ef1}), respectively, are coupled and nonlinear 
which  makes it quite difficult to find exact solutions.
However, as a first approximation it might be interesting to find the behaviour 
of the magnetic field neglecting the spatial dependence.
This is the long wavelength approximation. Considering variations over a 
characteristic comoving length scale $L$ much larger than the horizon $aH$
then the spatial derivatives of a quantity can be neglected with 
respect to its time derivatives (see for example, \cite{lwlap}).
In general, $\vec{E}$ and $\vec{B}$ can be written in terms of Fourier expansions,
\begin{eqnarray}
\vec{E}(\vec{x}, \eta)=\int d^3ke^{i\vec{k}\cdot\vec{x}}\vec{E}_k(\eta)
\hspace{2cm}
\vec{B}(\vec{x},\eta)=\int d^3ke^{i\vec{k}\cdot\vec{x}}\vec{B}_k(\eta).
\end{eqnarray}
Thus in the long wavelength approximation effectively only modes with small
wave numbers will contribute to the Fourier expansions.
Therefore, e.g. 
$\vec{B}(\vec{x},\eta)\simeq\int_0^{k_c}d^3k e^{i\vec{k}\cdot\vec{x}}\vec{B}_k(\eta)$.
Just using one mode 
$e^{i\vec{k}\cdot\vec{x}}\vec{B}_{k}(\eta)$ for  $k\ll k_c\stackrel{<}{\sim}
aH$ one can show that in the limit $k\rightarrow 0$ the terms involving 
spatial derivatives become subleading.
As it is commonly done, this approximation is applied to the second order equations
(\ref{b1}) and (\ref{ef1}) (cf., for example, \cite{tw}).

A  different way of looking at this is to use the stochastic approach to 
inflation where the mode expansion of a field, for example the inflaton, is 
separated into modes larger than the coarse graining domain and 
modes with wavelengths smaller than the coarse graining scale \cite{stoinf}.
The superhorizon modes contribute to the coarse grained field
which is made homogeneous by averaging over the coarse 
graining domain. The effect of modes leaving the coarse graining 
domain and contributing to the coarse grained field can effectively be 
modeled by a noise term in the equation of the coarse grained field.
Neglecting this backreaction effect the dynamics of the field 
on superhorizon scales is basically described by the homogeneous, 
coarse grained field.

Thus neglecting spatial derivatives equation (\ref{b1}) implies,
\begin{eqnarray}
\Bc''+\frac{L_X'}{L_X}\Bc'+\frac{L_Y'}{L_X}\Ec'
+\frac{L_Y'}{L_X}\left[\frac{L_X'}{L_X}\Ec-\frac{L_Y'}{L_X}\Bc
\right]
=0,
\label{b2}
\end{eqnarray} 
where $\Bc\equiv a^2\vec{B}_k$, $\Ec\equiv a^2\vec{E}_k$ 
and a prime denotes the derivative with respect to
conformal time $\eta$,  that is $' \equiv\frac{d}{d\eta}$.
In the case where the lagrangian only depends on $X$, $L_Y=0$, 
$\Bc=const.$ is a solution which corresponds to the 
conformally invariant case, that is linear electrodynamics.
In general, for $L_Y=0$,  equation (\ref{b2}) implies
\begin{eqnarray}
\Bc'=\frac{\vec{K}_k}{L_X},
\label{b3}
\end{eqnarray}
where $\vec{K}_k$ is a constant vector and $L_X\neq 0$. 
Moreover, linear electrodynamics is recovered for $\vec{K}_k\equiv 0$.

Furthermore, equation (\ref{ef1}) implies
\begin{eqnarray}
\frac{d}{d\eta}\left[
\Ec'+\frac{L_X'}{L_X}\Ec
-\frac{L_Y'}{L_X}\Bc\right]
\simeq 0.
\label{n1}
\end{eqnarray}
Equation (\ref{n1}) can be integrated to give
\begin{eqnarray}
\Ec'+\frac{L_X'}{L_X}\Ec
-\frac{L_Y'}{L_X}\Bc=\vec{P}_k,
\label{ec1}
\end{eqnarray}
where $\vec{P}_k$ is a constant vector.
The homogeneous part of equations (\ref{b1}) and (\ref{ef1})
are coupled non trivially because of $L_Y$, cf. equations (\ref{b2})
and (\ref{n1}).
Therefore in order to find solutions, the Lagrangian will be considered
to be  only a function of
$X$, $L=L(X)$.
Furthermore,  since $X=\frac{1}{2}(\vec{B}^2-\vec{E}^2)$ it is useful to find equations
for $\Ec^{\,2}$ and $\Bc^{\,2}$ which are given by, for $\vec{P}_k^2> 0$,
\begin{eqnarray}
\Ec^{\,2}\; ''+3\frac{L_X'}{L_X}\Ec^{\,2}\; '
+2\frac{L_X''}{L_X}\Ec^{\,2}&=&2\vec{P}^2_k
\label{E1}\\
\Bc^{\,2}\;  ''+\frac{L_X'}{L_X}\Bc^{\,2}\; '
-2\frac{\vec{K}^2_k}{L_X^2}&=&0.
\label{B1}
\end{eqnarray}
Assuming that the constant vector in equation (\ref{ec1}) vanishes, $\vec{P}_k=0$, leads to a significant  simplification. In this case, equation ({\ref{ec1}) for $L=L(X)$ can be solved immediately,
giving for the electric field
\begin{eqnarray}
\Ec=\frac{\vec{M}_k}{L_X},
\label{e1}
\end{eqnarray}
where $\vec{M}_k$ is a constant vector.
Thus for $\vec{P}_k=0$
equation (\ref{B1}) leads to an equation only
involving $X$ and $L_X$, namely,
\begin{eqnarray}
\frac{d^2}{d\eta^2}\left[2a^4X+\frac{\vec{M}^2_k}{L_X^2}\right]
+\frac{1}{L_X}\frac{dL_X}{d\eta}\frac{d}{d\eta}
\left[2a^4X+\frac{\vec{M}_k^2}{L_X^2}\right]-2\frac{\vec{K}_k^2}{L_X^2}=0.
\label{b4}
\end{eqnarray}

In order to solve this equation a particular lagrangian has to be chosen.
Here it is assumed that the lagrangian is of the 
form 
\begin{eqnarray}
L=-\left(\frac{X^2}{\Lambda^8}\right)^{\frac{\delta-1}{2}}X,
\label{L}
\end{eqnarray}
where $\delta$ is a dimensionless parameter and $\Lambda$ a dimensional
constant. This is the abelian Pagels-Tomboulis model \cite{asw}.
The nonabelian theory was proposed as an effective model of 
low energy QCD \cite{pt}. Evidently, linear electrodynamics is 
recovered for the choice $\delta=1$.
The lagrangian (\ref{L}) is chosen since it leads to a simplification of the equations,
but still allows to study the effects of a strongly nonlinear theory 
of electrodynamics on the generation of primordial magnetic fields.
In general, the energy-momentum tensor derived from a lagrangian 
$L(X)$ is given by
\begin{eqnarray}
T_{\mu\nu}=\frac{1}{4\pi}\left[L_Xg^{\alpha\beta}F_{\mu\alpha}
F_{\beta\nu}+g_{\mu\nu}L\right].
\label{T-PT}
\end{eqnarray}
Furthermore, for the lagrangian (\ref{L}) the trace of the energy-momentum tensor is given by
\begin{eqnarray}
T=\frac{1-\delta}{\pi}L,
\end{eqnarray}
which vanishes only in the case $\delta=1$ that is for 
linear electrodynamics.
In order to check if there are any constraints on the parameter $\delta$
the energy-momentum tensor is calculated explicitly.
The Maxwell tensor can be decomposed with respect to a fundamental
observer  with 4-velocity $u_{\mu}$ into an electric field $\vec{\hat{E}}$ and a magnetic field
$\vec{\hat{B}}$, following \cite{etb},
\begin{eqnarray}
F_{\mu\nu}=2\hat{E}_{[\mu}u_{\nu]}-\eta_{\mu\nu\kappa\sigma}u^{\kappa}\;
\hat{B}^{\sigma},
\label{Fmunu}
\end{eqnarray}
where $\eta_{\mu\nu\kappa\sigma}=\sqrt{-g}\epsilon_{\mu\nu\kappa\sigma}$ and 
$u_{\mu}u^{\mu}=-1$.
Then the electric and magnetic field are given, respectively, by
$\hat{E}_{\mu}=F_{\nu\mu}u^{\nu}$ and $\hat{B}_{\mu}=\frac{1}{2}\eta_{\mu\nu\omega\kappa}
u^{\nu}F^{\omega\kappa}$.  
The lab frame is defined by the proper lab coordinates $(t,\vec{r})$ determined by
$dt=ad\eta$, $d\vec{r}=ad\vec{x}$. Applying a coordinate transformation then 
gives the relation between the fields measured by a fundamental
observer and the lab frame. Thus using the four velocity of the fluid
$u^{\mu}=(a^{-1},0,0,0)$ this gives the relation  \cite{sb}
\begin{eqnarray}
E_i=a\hat{E}_i,\hspace{2cm} B_i=a\hat{B}_i.
\end{eqnarray}
As shown in \cite{etb} the energy-momentum tensor of an electromagnetic
field can be cast into the form of an imperfect fluid.
The energy-momentum tensor of an imperfect fluid is given by (see for example, \cite{etb}),
\begin{eqnarray}
T_{\mu\nu}=\rho u_{\mu}u_{\nu}+ph_{\mu\nu}+2q_{(\mu}u_{\nu)}+\pi_{\mu\nu},
\end{eqnarray}
where $\rho$ is the energy density, $p$ the pressure, $q_{\mu}$ the heat flux vector and 
$\pi_{\mu\nu}$ an anisotropic pressure contribution of the fluid. 
$h_{\mu\nu}=g_{\mu\nu}+u_{\mu}u_{\nu}$ is the metric on the space-like hypersurfaces 
orthogonal to $u_{\mu}$.
With $q_{\mu}u^{\mu}=0$ and $\pi_{\mu\nu}u^{\mu}=0$,
\begin{eqnarray}
\rho&=&T_{\mu\nu}u^{\mu}u^{\nu}\hspace{2cm}
q_{\alpha}=-T_{\mu\nu}u^{\mu}h^{\nu}_{\alpha}\nonumber\\
Q_{\alpha\beta}&\equiv &T_{\mu\nu}h^{\mu}_{\alpha}h^{\nu}_{\beta}
\hspace{2cm}
Q_{\alpha\beta}=ph_{\alpha\beta}+\pi_{\alpha\beta}.
\end{eqnarray}
Thus using equations (\ref{T-PT}) and (\ref{Fmunu})  the energy density and the heat flux vector for the Pagels-Tomboulis model (\ref{L}) are found to be
\begin{eqnarray}
\rho&=&-\frac{1}{8\pi}\frac{L}{X}\left[\left(2\delta-1\right)\hat{E}_{\alpha}
\hat{E}^{\alpha}+\hat{B}_{\alpha}
\hat{B}^{\alpha}\right]
\label{rho}\\
q_{\alpha}&=&\frac{\delta}{4\pi}\frac{L}{X}\eta_{\alpha\rho\xi\sigma}u^{\rho}
\hat{E}^{\xi}\hat{B}^{\sigma}.
\end{eqnarray}
Imposing the condition that $\pi_{\alpha\beta}$ is trace-free
then the pressure and $\pi_{\alpha\beta}$ are given by 
\begin{eqnarray}
p&=&\frac{1}{3}\rho-\frac{\delta-1}{3\pi}L\\
\pi_{\alpha\beta}&=&-\frac{\delta}{4\pi}\frac{L}{X}\left[
\frac{1}{3}\left(\hat{E}_{\mu}\hat{E}^{\mu}+\hat{B}_{\mu}\hat{B}^{\mu}\right)h_{\alpha\beta}
-\left(\hat{E}_{\alpha}\hat{E}_{\beta}
+\hat{B}_{\alpha}\hat{B}_{\beta}\right)\right].
\end{eqnarray}
Thus considering $\rho$ (cf. equation (\ref{rho})) in general there is a constraint 
on $\delta$.
Namely, the positivity of $\rho$ requires $\delta\geq\frac{1}{2}$.

Although in this work the abelian Pagels-Tomboulis model (cf. equation 
(\ref{L})) is used, for completeness,
other types of lagrangians describing theories of nonlinear
electrodynamics are briefly summarized in  the following.
The self-interaction energy of a slowly varying, but 
arbitrarily strong electromagnetic field was calculated by Heisenberg and 
Euler \cite{he, s}.  Expanding the resulting lagrangian into an 
asymptotic series gives a lagrangian of the form \cite{he}-\cite{bb}
\begin{eqnarray}
L=X+\kappa_0X^2+\kappa_1Y^2.
\end{eqnarray}
This describes the Heisenberg-Euler theory for
the choice $\kappa_0=\frac{8\alpha^2}{45m_e^4}$ and 
$\kappa_1=\frac{14\alpha^2}{45m_e^4}$, where $\alpha$ is the fine
structure constant and $m_e$ the electron mass.
Assuming the coefficients to be general and imagining a situation where 
the quadratic term in $X$ is dominant, the theory can be well approximated by the 
Pagels-Tomboulis lagrangian.

Born-Infeld theory is another example of a theory of nonlinear electrodynamics.
It was proposed as a classical, singularity-free theory of the electron \cite{bi}.
The lagrangian is given by (cf. \cite{bi-strings}-\cite{dbi})
\begin{eqnarray}
L=\frac{1}{\beta}\left(1-\sqrt{1+2\beta^2X-\beta^4Y^2}
\right),
\end{eqnarray}
where $\beta$ is a parameter.  This type of action also appears 
in the description of open string states in string or M theory  \cite{bi-strings}-\cite{dbi}.
In this case $\beta=2\pi\alpha'$ with  $\alpha'$  the string tension.
Considering a general parameter $\beta$ and moreover the case in 
which  the term $\beta^2X$ is dominant
results in a lagrangian of the Pagels-Tomboulis form.
Furthermore, the parameter $\delta$ in (\ref{L}) is given by
$\delta=\frac{1}{2}$.

However, since both invariants $X$ and $Y$ appear, the resulting 
equations are non trivially coupled in $X$ and $Y$ which makes it 
difficult to find solutions in closed form.
In order to study the effects of nonlinear electrodynamics in the 
early universe, the abelian Pagels-Tomboulis theory (cf. equation (\ref{L}))
will be used. This has the advantage that even though it is a strongly
nonlinear theory of electrodynamics, it is still possible to find 
approximate solutions in certain regimes.

Therefore, in the following we will assume that the theory is determined
by the abelian Pagels-Tomboulis lagrangian (\ref{L}).

\section{Estimating the magnetic field strength in the Pagels-Tomboulis model}
\setcounter{equation}{0}

The following model will be considered. During de Sitter inflation electrodynamics is
nonlinear and described by the Pagels-Tomboulis lagrangian (\ref{L}). This means that 
electrodynamics is highly nonlinear and very different from standard Maxwell
electrodynamics.
At the end of inflation electrodynamics becomes linear and thus the description
of reheating and the subsequent radiation dominated stage is unaltered.

The scale factor during de Sitter is given by
\begin{eqnarray}
a(\eta)=a_1\left(\frac{\eta}{\eta_1}\right)^{-1},
\label{A}
\end{eqnarray}
where $\eta\leq\eta_1<0$.
The end of inflation is assumed to be at $\eta=\eta_1$.

Equations (\ref{E1}) and (\ref{B1}) are coupled, since $X$ depends on $\vec{E}^2$ and 
$\vec{B}^2$, in particular the invariant $X$ reads,
$2a^4X\simeq\Bc^{\,2}-\Ec^{\,2}$.
In order to make progress three different regimes of approximation will be considered.
\begin{enumerate}
\item $\Bc^{\,2}\simeq{\cal O}(\Ec^{\,2})$.
\item $\Bc^{\,2}\gg \Ec^{\,2}$. This implies the approximation
$2 a^4X\simeq \Bc^{\,2}$.
\item $\Ec^{\,2}\gg \Bc^{\,2}$. This implies the approximation
$2 a^4X\simeq -\Ec^{\,2}$.
\end{enumerate}

As will be explained further on, following \cite{tw} it is assumed that 
quantum fluctuations in the electromagnetic field provide an 
initial magnetic and electric field. Therefore, it seems 
naturally to expect that initially $\Bc^2\simeq{\cal O}(\Ec^2)$.
Thus the three cases mentioned above correspond to different types of 
evolution of the ratio $\Bc^2/\Ec^2$ during inflation.
After the end of inflation, during the radiation era, 
while the electric field decays 
rapidly due to plasma effects, the magnetic field remains 
(see, e.g., \cite{tw,d93}).

\subsection{Case {\it i.)} $\Bc^{\,2}\simeq{\cal O}(\Ec^{\,2})$ }
 \label{case1}
 
As a  further simplification it is assumed that the contribution of the constant vector $\vec{P}_k^2$ in 
equation (\ref{E1}) vanishes. 
In this case the equation (\ref{b4}) during inflation in the Pagels-Tomboulis model (\ref{L})
can be written in the form 
\begin{eqnarray}
\left[\left(\frac{x}{x_1}\right)^{-4}-\alpha_1(\delta-1)y^{-2\delta+1}\right]\ddot{y}&=&
-(\delta-1)\left[\alpha_1\delta y^{-2\delta+1}+\left(\frac{x}{x_1}\right)^{-4}\right]
\frac{\dot{y}^2}{y}\nonumber\\
&&+\frac{4(\delta+1)}{x_1}\left(\frac{x}{x_1}\right)^{-5}\dot{y}
-\frac{20}{x_1^2}\left(\frac{x}{x_1}\right)^{-6}y
\nonumber\\
&&+\alpha_2y^{-2(\delta-1)},
\label{fode}
\end{eqnarray}
where a dot denotes the derivative with repect to $x$ and $y=y(x)$. Furthermore the following
definitions have been used,
\begin{eqnarray}
x=\frac{\eta}{M_P^{-1}} ,\hspace{1cm}
y\equiv\frac{X}{\Lambda^4} ,\hspace{1cm}
\alpha_1\equiv\frac{\hat{M}^2}{\hat{\Lambda}^4\delta^2a_1^4}, \hspace{1cm}
\alpha_2\equiv\frac{\hat{K}^2}{\hat{\Lambda}^4\delta^2a_1^4}
\label{const}
\end{eqnarray}
Moreover, the hatted quantities are dimensionless constants, 
\begin{eqnarray}
\hat{\Lambda}\equiv\frac{\Lambda}{M_P}, \hspace{1cm}
\hat{M^2}\equiv\frac{\vec{M}_k^2}{M_P^4}, \hspace{1cm}
\hat{K}^2\equiv\frac{\vec{K}_k^2}{M_P^6}
\end{eqnarray}
Finally,
$M_P$ is the Planck mass.
Equation (\ref{fode}) is a nonlinear differential equation and thus to find 
exact solutions is not trivial. Therefore, in order to proceed one further approximation will be made.
It turns out that there is an approximate solution in closed form for $\delta>1$. In this case
neglecting the terms involving $\left(\frac{x}{x_1}\right)^{\alpha}$, with the exponents
$\alpha=-4,-5,-6$,  yields the equation 
\begin{eqnarray}
y\ddot{y}=\delta(\dot{y})^2+\frac{1}{1-\delta}m^2y^2,
\label{apode}
\end{eqnarray}
where $m^2\equiv\frac{\alpha_2}{\alpha_1}$
which is solved by
\begin{eqnarray}
y(x)=C_2\left[\cosh\left(mx+(\delta-1)mC_1\right)\right]^{\frac{1}{1-\delta}},
\label{b5}
\end{eqnarray}
where $C_1$, $C_2$ are constants.
In general, the magnetic field is given by
\begin{eqnarray}
\vec{B}^2_k=2X+\frac{\vec{M}_k^2}{\delta^2a^4}\left(\frac{X}{\Lambda^4}\right)^{-2(\delta-1)}.
\end{eqnarray}
However, for the approximate solution (\ref{b5}) the first term becomes subdominant and 
the magnetic field can be approximated by
\begin{eqnarray}
\vec{B}^2_k\simeq\frac{\vec{M}_k^2}{\delta^2a^4}\left(\frac{X}{\Lambda^4}\right)^{-2(\delta-1)}.
\label{B-app}
\end{eqnarray}
Thus, the magnetic field at the end of inflation at the time $\eta_1$ can be expressed 
in terms of the magnetic field at the time, say $\eta_2$, when the comoving length 
scale $\lambda$ was crossing the horizon during inflation.
Thus  
\begin{eqnarray}
\frac{B^2_k(a_1)}{B^2_k(a_2)}\simeq e^{-4N(\lambda)}\frac{\cosh^2[m(x_1+(\delta-1)C_1)]}{\cosh^2[m(x_2+(\delta-1)C_1)]},
\end{eqnarray}
where $N(\lambda)$ is the number of e-folds before the end of inflation 
at which $\lambda$ left the horizon, that is, $e^{N(\lambda)}=a_1/a_2$.
Furthermore, the constant $C_1$ is chosen such that 
$(\delta-1)C_1=-x_2$. Using that during de Sitter inflation, $a=a_1(\eta_1/\eta)$ 
and the number of e-folds, results in the magnetic energy density $\rho_B$ at the end of inflation,
\begin{eqnarray}
\rho_B(a_1)\simeq\rho_B(a_2)e^{-4N(\lambda)}\cosh^2[-mx_1(e^{N(\lambda)}-1)],
\end{eqnarray}
where $\rho_B=\frac{B^2}{8\pi}$. Following \cite{tw} the 
ratio of magnetic energy density to radiation energy density, $r$ is 
introduced,
\begin{eqnarray}
r\equiv\frac{\rho_B}{\rho_{\gamma}}.
\end{eqnarray}
In the case of linear electrodynamics, the energy density in the magnetic 
field and the radiation background decay both as $a^{-4}$ and thus the ratio
$r$ stays invariant as the universe expands.
In order to seed a galactic dynamo $r$ has to be at least $r=10^{-37}$ corresponding to a magnetic seed field at the time of galaxy formation $B_s\simeq 10^{-20}$G. In order to seed the galactic field directly, without a galactic
dynamo operating, $r$ has to be of order $r=10^{-8}$.
Furthermore, we also note, that in a flat universe with a cosmological 
constant, these bounds can be lowered significantly. In this case, $r$ has to be at least
$r=10^{-57}$  to successfully seed a galactic dynamo \cite{dlt}. This implies that 
the magnetic field at the time of galaxy formation has to be at least of order
$B_s\simeq 10^{-30}$ G.
Following \cite{tw}, it is assumed that the energy density stored in the
mode with comoving wavelength $\lambda$ is of the order of the 
energy density in a thermal bath at the Gibbons-Hawking temperature
of de Sitter space.
This leads to $\rho_B(a_2)\simeq H^4\simeq \left(\frac{M^4}{M_p^2}\right)^2$,
where the constant energy density during inflation is given by
$M^4$. Finally, using that at the end of inflation the energy density 
in the radiation background is given by $\rho_{\gamma}=M^2T_{RH}^2$, where
$T_{RH}$ is the reheat temperature, the ratio of magnetic to radiation 
energy density at the end of inflation at $\eta=\eta_1$ is found to be,
\begin{eqnarray}
r(a_1)\simeq\left(\frac{M}{M_P}\right)^6\left(\frac{T_{RH}}{M_P}\right)^{-2}
e^{-4N(\lambda)}\cosh^2\left[-mx_1\left(e^{N(\lambda)}-1\right)\right].
\end{eqnarray}
Furthermore, following \cite{tw}, the number of e-folds can be found as
\begin{eqnarray}
e^{N(\lambda)}\simeq 9.2\times10^{25}\left(\frac{\lambda}{{\rm Mpc}}\right)
\left(\frac{M}{M_P}\right)^{\frac{2}{3}}\left(\frac{T_{RH}}{M_P}\right)^{\frac{1}{3}},
\end{eqnarray}
where it is assumed that the scale factor today is $a_0=1$ and thus
comoving and physical scales coincide in the present.
Thus the ratio $r$ at the end of inflation is given by
\begin{eqnarray}
r(a_1)\simeq 10^{-104}\left(\frac{\lambda}{\rm Mpc}\right)^{-4}
\left(\frac{M}{T_{RH}}\right)^{\frac{10}{3}}
\cosh^2\left[-9.2\times10^{25}\left(\frac{\lambda}{\rm Mpc}\right)
\left(\frac{M}{M_P}\right)^{\frac{2}{3}}\left(\frac{T_{RH}}{M_P}
\right)^{\frac{1}{3}}mx_1\right].
\label{r1}
\end{eqnarray}
Therefore, in order to achieve, a ratio of magnetic to radiation energy density
at the end of inflation, which is at least some value $r_0$, that is 
$r(a_1)\geq r_0$, $mx_1$ has to satisfy,
\begin{eqnarray}
-mx_1\geq 10^{-26}\left(\frac{\lambda}{\rm Mpc}\right)^{-1}
\left(\frac{M}{M_P}\right)^{-\frac{2}{3}}
\left(\frac{T_{RH}}{M_P}\right)^{-\frac{1}{3}}
{\rm arccosh}\left[10^{52}\left(\frac{\lambda}{\rm Mpc}\right)^2
\left(\frac{T_{RH}}{M}\right)^{\frac{5}{3}}r_0^{\frac{1}{2}}\right].
\label{e348}
\end{eqnarray}
Equation (\ref{e348}) only implies a constraint on $mx_1$ if the 
argument of arccosh is bigger or equal to one which can be 
interpreted as a bound on the reheat temperature.
Assuming a galactic scale, that is $\lambda=1$Mpc implies,
\begin{eqnarray}
10^{52}\left(\frac{T_{RH}}{M}\right)^{\frac{5}{3}}r_0^{\frac{1}{2}}\geq 1,
\end{eqnarray}
which for $r_0=10^{-37}$ implies $T_{RH}\geq 1$ MeV, where it was assumed \cite{tw} that the inflationary energy scale is given by
$M=10^{17}$GeV. This is always satisfied since the reheat temperature 
has to be at least 10 MeV (for even lower values, see \cite{hanne}) in order to allow for nucleosynthesis 
to take place unaltered. For a smaller value of $r_0$, say $r_0=10^{-57}$,
the reheat temperature is required to be at least of order $10^3$ GeV.
However, in this work the reheat temperature is assumed to be
at least $T_{RH}\geq 10^9$ GeV.
There is also an upper bound on $-mx_1$ coming from the requirement that 
$r<1$ in order not to overclose the universe.
This implies,
\begin{eqnarray}
-mx_1<10^{-26}\left(\frac{\lambda}{\rm Mpc}\right)^{-1}
\left(\frac{M}{M_P}\right)^{-\frac{2}{3}}
\left(\frac{T_{RH}}{M_P}\right)^{-\frac{1}{3}}
{\rm arccosh}\left[10^{52}\left(\frac{\lambda}{\rm Mpc}\right)^2
\left(\frac{T_{RH}}{M}\right)^{\frac{5}{3}}\right].
\label{e350}
\end{eqnarray}
The constraint equations (\ref{e348}) and (\ref{e350}) can always
be satisfied, since, by assumption, $r_0\leq 1$ and, moreover, for
physically interesting models $r_0\ll 1$. Thus assuming $\lambda=1$ Mpc
and $M=10^{17}$ GeV,
the following values for $-mx_1$ are found.
For a model with reheat temperature $T_{RH}=10^9$ GeV \cite{tw}
the parameter $-mx_1$ has to be in the range $2.7\times 10^{-20}<-mx_1<5\times 10^{-20}$
in order to achieve a magnetic seed field with a field strength to be 
at least $B_s\simeq 10^{-20}$G, 
corresponding to $r_0=10^{-37}$. For a higher reheat temperature $T_{RH}=10^{17}$GeV \cite{tw},
for the same magnetic seed field strength $-mx_1$ has to be in the range,
$9.5\times 10^{-23}<-mx_1<1.5\times 10^{-22}$.
And similarly, for the less conservative bound $r_0=10^{-57}$, for 
$T_{RH}=10^9$ GeV, $-mx_1$ has to be in the range
$1.4\times 10^{-20}<-mx_1<5\times 10^{-20}$ and for
$T_{RH}=10^{17}$ GeV it is found that 
$6.7\times 10^{-23}<-mx_1<1.5\times 10^{-22}$.

Thus there is a range of parameters for which strong enough magnetic seed
fields can be created in the Pagels-Tomboulis model of nonlinear 
electrodynamics.
Since the analysis is based on the approximate exact solution given by
equation (\ref{b5}) it is also important to check that the solution 
provides a good approximation to the solution of the full differential
equation (\ref{fode}). This has been done in Appendix A.

In summary, for $\delta>1$, there is an approximate analytical solution 
which allows to find an expression for the ratio of the magnetic to radiation
energy density. There is a range of parameters for which 
magnetic seed fields of cosmologically interesting field strengths can be created.

\subsection{Case {\it ii.)}   $\Bc^{\,2}\gg \Ec^{\,2}$ }

In this case equation (\ref{B1}) takes the form,
\begin{eqnarray}
\frac{d^2}{d\eta^2}\left(a^4 X\right)+(\delta-1)\frac{1}{X}
\frac{dX}{d\eta}\frac{d}{d\eta}
\left(a^4 X\right)-\frac{\vec{K}^2_k}{\delta^2}\left(\frac{X}{\Lambda^4}\right)^{-2(\delta-1)}=0,
\label{eq351}
\end{eqnarray}
where it has been used that $2X\simeq\vec{B}_k^2$.
It is possible to find different types of solutions of equation (\ref{eq351}) depending on the value of 
the parameter $\delta$ of the Pagels-Tomboulis model. On the one hand  there are power law solutions for $\delta\neq\frac{1}{2}$ and $\delta\neq\frac{5}{4}$. On the 
other hand there are solutions with a distinct behaviour for $\delta=\frac{1}{2}$ and $\delta=\frac{5}{4}$.
Actually of the latter ones only the case $\delta=\frac{1}{2}$ will be discussed explicitly.
This is so because for $\delta=\frac{5}{4}$ it is only possible to find an implicit solution depending 
on Euler's $\beta$ function which makes it very difficult to estimate the primordial
magnetic field strength.

\subsubsection{Solution for $\delta\neq\frac{1}{2}$ and $\delta\neq\frac{5}{4}$}

For $\delta\neq\frac{1}{2}$ and $\delta\neq\frac{5}{4}$,  equation (\ref{eq351}) can be solved by a power-law function,
\begin{eqnarray}
X=X_1\left(\frac{\eta}{\eta_1}\right)^{\alpha}.
\end{eqnarray}
This leads to the solution for the magnetic field
\begin{eqnarray}
\vec{B}^2_k=2\Lambda^4\left[\frac{2}{\alpha_2x_1^2}\left(\frac{5-4\delta}{2\delta-1}\right)^2
\right]^{\frac{1}{1-2\delta}}\left(\frac{\eta}{\eta_1}\right)^{\frac{6}{2\delta-1}},
\label{bb1}
\end{eqnarray}
where, as before,  $\alpha_2\equiv\frac{\hat{K}^2}{a_1^4\hat{\Lambda}^4\delta^2}$.
Thus using the definitions as given for case {\it i.)}  (cf. section \ref{case1})
the ratio of magnetic energy density
to radiation energy density at the end of inflation $r(a_1)$ is found to be, for $\delta\neq\frac{1}{2}$, 
$\delta\neq\frac{5}{4}$,
\begin{eqnarray}
r(a_1)\simeq \left(9.2\times 10^{25}\right)^{-\frac{6}{2\delta-1}}
\left(\frac{\lambda}{\rm Mpc}\right)^{-\frac{6}{2\delta-1}}\left(
\frac{M}{M_P}\right)^{2\frac{6\delta-5}{2\delta-1}}
\left(\frac{T_{RH}}{M_P}\right)^{-\frac{4\delta}{2\delta-1}}
\label{r_1}
\end{eqnarray}

The range of validity of the assumption $\Bc^{\,2}\gg \Ec^{\,2}$ can be checked
to first order by using the solution for $B^2_k\simeq 2X$, (cf. equation (\ref{bb1})) in the equation
for $\Ec^{\,2}$, equation (\ref{E1}).
This leads to
\begin{eqnarray}
\Ec^{\,2}\; ''+\beta_1\eta^{-1}\Ec^{\,2}\; '
+\beta_2\eta^{-2}\Ec^{\, 2}=2\vec{P}_k^2,
\end{eqnarray}
where $\beta_1\equiv 3\alpha(\delta-1)$ and $\beta_2\equiv 2\alpha(\delta-1)
\left[\alpha\left(\delta-1\right)-1\right]$.
This equation is solved by, for $\vec{P}_k^2\neq 0$,
\begin{eqnarray}
\Ec^{\,2}=\vec{P}_k^2\left(\frac{2\delta-1}{8\delta-7}\right)^2\eta^2
+c_0\left(\frac{\eta}{\eta_1}\right)^{\frac{-4\delta+5}{2\delta-1}}
+c_1\left(\frac{\eta}{\eta_1}\right)^{-\frac{12(\delta-1)}{2\delta-1}},
\label{ecase2}
\end{eqnarray}
where $c_0$ and $c_1$ are constants. For $\vec{P}_k^2=0$, the solution is given by 
equation (\ref{e1}) which will be discussed below.
During de Sitter inflation the scale factor is given by equation (\ref{A}}).
Thus, finally, the ratio $\Ec^{\,2}/\Bc^{\,2}$ is given by
\begin{eqnarray}
\frac{\Ec^{\,2}}{\Bc^{\,2}}\simeq \mu_0\left(\frac{\eta}{\eta_1}\right)^{12\frac{\delta-1}{2\delta-1}}
+\mu_1\left(\frac{\eta}{\eta_1}\right)^{\frac{4\delta-5}{2\delta-1}}
+\mu_2\left(\frac{\eta}{\eta_1}\right)^{-2},
\end{eqnarray}
where $\mu_0$, $\mu_1$ and $\mu_2$ are constants depending on 
the constants of the solutions of the electric and magnetic field.
However, their explicit form is not important here.
Since during inflation, $\eta<\eta_1<0$, and hence $\eta/\eta_1>1$.
At $\eta=\eta_2$, that is at the time when the mode is leaving the
horizon during inflation, the initial condition,
$\frac{\Ec^2}{\Bc^2}(\eta_2)=1$ is imposed.
In order for the approximation to be consistent, it is 
required that the solution evolves such that
$\frac{\Ec^2}{\Bc^2}\leq 1$. So assuming that each of the terms
is of the order of $\frac{1}{3}$ at $\eta=\eta_2$ then the constants
$\mu_0$, $\mu_1$ and $\mu_2$ can be estimated in terms of 
$\frac{\eta_2}{\eta_1}$. This leads to 
\begin{eqnarray}
\frac{\Ec^2}{\Bc^2}=\frac{1}{3}\left(\frac{\eta}{\eta_2}\right)^{12
\frac{\delta-1}{2\delta-1}}+\frac{1}{3}\left(\frac{\eta}{\eta_2}\right)^
{\frac{4\delta-5}{2\delta-1}}+\frac{1}{3}\left(\frac{\eta}{\eta_2}\right)
^{-2}.
\label{e-b}
\end{eqnarray}
Thus the last term is growing and in general, $\frac{\Ec^2}{\Bc^2}$ 
is not smaller than one. Thus in order for the solution to be consistent
within the approximation, the constant $c_1$ in equation (\ref{ecase2}) has to be set to zero. Furthermore, the exponents in equation (\ref{e-b}) have to be positive,
imposing the constraints, $\delta<\frac{1}{2}$ or $\delta>\frac{5}{4}$.
The former one is ruled out since $\delta\geq\frac{1}{2}$.
Then by assuming that the two remaining terms, after setting $c_1$ to zero, 
contribute equally at $\eta=\eta_2$, the ratio $\frac{\Ec^2}{\Bc^2}\leq 1$ for $\eta\geq\eta_2$ and $\delta>\frac{5}{4}$.

In order to seed the galactic dynamo, it is required that $r\geq r_0$ where $r_0$ is the lower bound on the strength of the magnetic field.
In the expression for $r=\frac{\rho_B}{\rho_{\gamma}}$ at the end of inflation 
there are, apart from $\delta$, two parameters: on the one hand the constant
energy density during inflation given by $M^4$ and on the other hand the 
reheat temperature, $T_{RH}$. Following \cite{tw} $M$ is chosen to be 
$M=10^{17}$ GeV. 
The reheat temperature depends on the details of the reheating process.
It can be as low as $4$ MeV  \cite{hanne} and in general, one expects an upper limit 
$T_{RH}\leq M$. However, in supersymmetric theories this limit 
is lowered down to $10^9$ GeV \cite{bebe}.

In Figure 1 $\log r$ is plotted against the Pagels-Tomboulis
parameter $\delta$ for the inflationary energy scale $M=10^{17}$ GeV for the 
reheat temperatures $T_{RH}=10^{17}$ GeV  and $T_{RH}=10^9$ GeV.
\begin{figure}[ht]
\centerline{\epsfxsize=3in\epsfbox{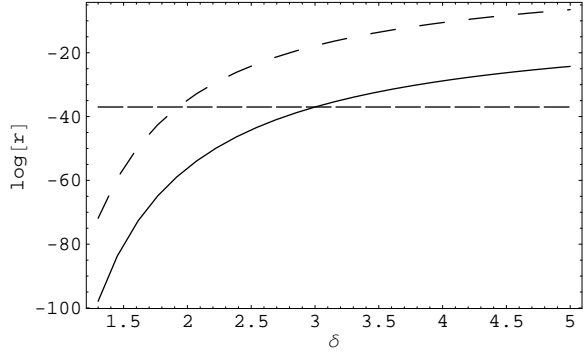}}
\caption{For $\lambda=1$ Mpc $\log r$ (cf. equation (\ref{r_1}))  is shown as a function 
of $\delta$ for $T_{RH}=10^{17}$ GeV (black line) and $T_{RH}=10^{9}$ GeV (long-dashed line).
The dashed line corresponds to $r=10^{-37}$.}
\label{fig}
\end{figure}
As can be seen from Figure 1 
for $\delta>\frac{5}{4}$ there is range of 
$\delta$ for which primordial magnetic fields with cosmologically interesting
field strengths can be generated.

In the case $\vec{P}_k^2=0$ the solution for $\Ec$ is 
given by equation (\ref{e1}).  This leads to
\begin{eqnarray}
\Ec^{\,2}\simeq\frac{\vec{M}_k^{\,2}}{\delta^2}
\left(\frac{X^2}{\Lambda^8}\right)^{1-\delta}
\end{eqnarray}
and thus
\begin{eqnarray}
\frac{\Ec^{\,2}}{\Bc^{\,2}}\simeq\frac{1}{m^2x_1^2}\left(\frac{5-4\delta}{2\delta-1}\right)^2
\left(\frac{\eta}{\eta_1}\right)^{-2},
\end{eqnarray}
where as before (cf. equation (\ref{fode})) $m^2=\frac{\alpha_2}{\alpha_1}$.
Thus imposing the initial condition $\frac{\Ec^2}{\Bc^2}(\eta_2)\simeq 1$
implies 
\begin{eqnarray}
\frac{\Ec^2}{\Bc^2}=\left(\frac{\eta_2}{\eta}\right)^2,
\end{eqnarray}
which implies $\frac{\Ec^2}{\Bc^2}\geq 1$ for $\eta\geq\eta_2$. Thus 
this solution is not consistent with the approximation.

\subsubsection{Solution for $\delta=\frac{1}{2}$}

In this case equation (\ref{eq351}) can be written as
\begin{eqnarray}
\ddot{g}-\frac{1}{2}\left(\frac{4}{x}+\frac{\dot{g}}{g}\right)\dot{g}-\alpha_2\left(\frac{x}{x_1}\right)^4g=0,
\end{eqnarray}
where $g\equiv \left(\frac{x}{x_1}\right)^{-4}X$, $x\equiv\eta/M_P^{-1}$ and $\alpha_2$ a constant as defined before, (cf. equation (\ref{const})).
This is solved by $g(x)=c_2\cosh^2\left[\left(\frac{\alpha_2}{18}\right)^{\frac{1}{2}}x_1\left(
\frac{x}{x_1}\right)^3+c_1\right]$, where $c_1$ and $c_2$ are constants.
Thus with $\vec{B}^2_k\simeq 2X$ it follows that 
\begin{eqnarray}
\vec{B}^2_k\simeq 2c_2\left(\frac{x}{x_1}\right)^4\cosh^2\left[\left(\frac{\alpha_2}{18}\right)^{\frac{1}{2}}x_1\left(\frac{x}{x_1}\right)^3+c_1\right].
\label{B0.5}
\end{eqnarray}
This then leads to the ratio of magnetic energy density to radiation background energy density
$r=\frac{\rho_B}{\rho_{\gamma}}$ at the end of inflation,
\begin{eqnarray}
r(a_1)\simeq 10^{-104}\left(\frac{\lambda}{{\rm Mpc}}\right)^{-4}
\left(\frac{M}{T_{RH}}\right)^{\frac{10}{3}}\cosh^2\left[
-8\times 10^{77}x_1\left(\frac{\alpha_2}{18}\right)^{\frac{1}{2}}\left(\frac{\lambda}{{\rm Mpc}}\right)^3
\left(\frac{M}{M_P}\right)^2\frac{T_{RH}}{M_P}\right].
\label{r_1_2}
\end{eqnarray}
Here the constant $c_1$ has been chosen as $c_1\equiv-\left(\frac{\alpha_2}{18}\right)^{\frac{1}{2}}
x_1\left(\frac{\eta_2}{\eta_1}\right)^3$.
Imposing the condition that $r_0<r(a_1)<1$ results, for $\lambda=1$ Mpc, in 
\begin{eqnarray}
10^{-78}\left(\frac{M}{M_P}\right)^{-2}\left(\frac{T_{RH}}{M_P}\right)^{-1}
{\rm arccosh}\left[10^{52}\left(\frac{T_{RH}}{M}\right)^{\frac{5}{3}}r_0^{\frac{1}{2}}\right]<
-x_1\left(\frac{\alpha_2}{18}\right)^{\frac{1}{2}}
\nonumber\\
<
10^{-78}\left(\frac{M}{M_P}\right)^{-2}\left(\frac{T_{RH}}{M_P}\right)^{-1}
{\rm arccosh}\left[10^{52}\left(\frac{T_{RH}}{M}\right)^{\frac{5}{3}}\right].
\label{bounds}
\end{eqnarray}
The resulting values for different choices of $T_{RH}$, $M$ and $r_0$ are given in Table 1.
\begin{table}[ht]
\begin{center}
\begin{tabular}{|c|c|c||c|c|}
\hline
$T_{RH}$(GeV) & $M$(GeV) &  $r_0$ & $-x_1\left(\frac{\alpha_2}{18}\right)^{\frac{1}{2}}_{low} $&
$ -x_1\left(\frac{\alpha_2}{18}\right)^{\frac{1}{2}}_{up}$
\\
\hline \hline
$10^{9}$  & $10^{17}$  & $10^{-37}$ & $8.6\times 10^{-63}$ & $1.6\times 10^{-62}$\\
\hline
$10^9$ &  $10^{17}$  & $10^{-57}$ & $4.4\times 10^{-63}$ & $1.6\times 10^{-62}$\\
\hline
$10^{17}$  & $10^{17}$  & $10^{-37}$ & $8.6\times 10^{-71}$ & $1.6\times 10^{-70}$\\
\hline
$10^{17}$  & $10^{17}$  & $10^{-57}$ & $4.4\times 10^{-71}$ & $1.6\times 10^{-70}$\\
\hline
\end{tabular}
\caption{Lower and upper bounds of $-x_1\left(\frac{\alpha_2}{18}\right)^{\frac{1}{2}}$ 
derived from equation (\ref{bounds}) for different values of the reheat temperature $T_{RH}$,
the constant energy density during inflation determined by $M$ and the lower limit 
on the field strength of a primordial magnetic seed field determined by $r_0$.
The notation used indicates $-x_1\left(\frac{\alpha_2}{18}\right)^{\frac{1}{2}}_{low}<
-x_1\left(\frac{\alpha_2}{18}\right)^{\frac{1}{2}}<-x_1\left(\frac{\alpha_2}{18}\right)^{\frac{1}{2}}_{up}$.}
\end{center}
\end{table}
Furthermore, in order to check the validity of the solution (\ref{B0.5}) which was derived under the assumption that $\Ec^{\,2}/\Bc^{\,2}\ll 1$, we consider the cases $\vec{P}_k^2>0$ and 
$\vec{P}_k^2=0$.
In the case $\vec{P_k^2}>0$ the electric field strength is determined by
equation (\ref{E1}). As a first approximation, the solution for $X\simeq\frac{1}{2}\vec{B}_k^2$,
where $\vec{B}_k^2$ is given by (\ref{B0.5}), will be used in (\ref{E1}). For consistency, the resulting solution for the electric field strength 
should be much smaller than the magnetic field strength.
In equation (\ref{E1}) the expressions for $\frac{L_X'}{L_X}$ and $\frac{L_X''}{L_X}$ are needed which are given in Appendix B.
The cosmologically interesting values of $\mu\equiv-\left(\frac{\alpha_2}{18}\right)^{\frac{1}{2}}x_1$
are very small, $\mu\lesssim{\cal O}(10^{-62})$, as can be seen from Table 1.
Thus to zeroth order in $\mu$ equation (\ref{E1}) becomes,
\begin{eqnarray}
\Ec^{\,2}\; ''-\frac{6}{\eta}\Ec^{\,2}\;'+\frac{12}{\eta^2}\Ec^{\,2}=2\vec{P}_k^2,
\end{eqnarray}
which is solved by 
\begin{eqnarray}
\Ec^{\,2}=\vec{P}_k^2\eta_1^2\left(\frac{\eta}{\eta_1}\right)^2+\beta_0\left(\frac{\eta}{\eta_1}
\right)^3+\beta_1\left(\frac{\eta}{\eta_1}\right)^4,
\end{eqnarray}
where $\beta_0$ and $\beta_1$ are constants.
Therefore
\begin{eqnarray}
\frac{\Ec^{\,2}}{\Bc^{\,2}}\simeq\frac{\vec{P}_k^2\eta_1^2\left(\frac{\eta}{\eta_1}\right)^2+\beta_0\left(\frac{\eta}{\eta_1}\right)^3+\beta_1\left(\frac{\eta}{\eta_1}\right)^4}
{2c_2a_1^4\cosh^2\left[-\left(\frac{\alpha_2}{18}\right)^{\frac{1}{2}}x_1\left[\left(\frac{\eta_2}{\eta_1}\right)^3-
\left(\frac{\eta}{\eta_1}\right)^3\right]\right]}.
\end{eqnarray}
Imposing the initial condition $\frac{\Ec^2}{\Bc^2}(\eta_2)=1$ it can be seen that
$\frac{\Ec^{\,2}}{\Bc^{\,2}}$ is decreasing very fast and thus the solution (\ref{B0.5}) is consistent at lowest order in $\mu$.

In the case $\vec{P}_k^2=0$ the solution for the electric field strength is given by
equation (\ref{e1}), leading to 
\begin{eqnarray}
\Ec^{\,2}\simeq 4 \vec{M}_k^2\left(\frac{X^2}{\Lambda^8}\right)^{\frac{1}{2}}.
\end{eqnarray}
This implies 
\begin{eqnarray}
\frac{\Ec^{\,2}}{\Bc^{\,2}}\simeq\frac{\alpha_1}{2}\left(\frac{\eta}{\eta_1}\right)^4.
\end{eqnarray}
Imposing that initially $\frac{\Ec^2}{\Bc^2}(\eta_2)=1$ it follows that 
\begin{eqnarray}
\frac{\Ec^{\,2}}{\Bc^{\,2}}=\left(\frac{\eta}{\eta_2}\right)^4,
\end{eqnarray}
which is smaller than one for $\eta>\eta_2$. Hence the solution is
consistent with the approximation.

Thus, for the solution in this case, there is a choice of parameters which 
allows to create primordial magnetic fields of cosmologically interesting 
field strengths. This holds for both cases, $\vec{P}^2>0$ and $\vec{P}^2=0$.

\subsection{Case {\it iii.)}   $\Ec^{\,2}\gg \Bc^{\,2}$ }

Starting out at the same order of magnitude, 
in this approximation at the end of inflation the energy density in the 
electric field is much larger than in the magnetic field.
The approximation implies that $\Ec^{\,2}\simeq-2a^4X$.
In the case $\vec{P}_k^2>0$
equation (\ref{E1}) yields to
\begin{eqnarray}
\frac{d^2}{d\eta^2}\left(a^4X\right)+3(\delta-1)\frac{X'}{X}
\frac{d}{d\eta}\left(a^4X\right)+2(\delta-1)\left[\frac{X''}{X}+
(\delta-2)\left(\frac{X'}{X}\right)^2\right]a^4X=-\vec{P}^2_k,
\end{eqnarray}
which is solved by, for $\delta\neq\frac{5}{6}$,
\begin{eqnarray}
X=-\frac{\vec{P}_k^2\eta_1^2}{2a_1^4(6\delta-5)^2}
\left(\frac{\eta}{\eta_1}\right)^6.
\label{solx}
\end{eqnarray}
Thus using $X=X_1\left(\frac{\eta}{\eta_1}\right)^{\alpha}$ in the equation
determining the magnetic field (cf. equation (\ref{B1})) gives,
\begin{eqnarray}
\vec{B}_k^2=\frac{\vec{K}_k^2\eta_1^2}
{a_1^4\delta^2\left[1-\alpha(\delta-1)\right]^2}
\left(\frac{X_1^2}{\Lambda^8}\right)^{-(\delta-1)}
\left(\frac{\eta}{\eta_1}\right)^{6-2\alpha(\delta-1)}
+b_0\left(\frac{\eta}{\eta_1}\right)^4
+b_1\left(\frac{\eta}{\eta_1}\right)^{5-\alpha(\delta-1)},
\label{e341}
\end{eqnarray} 
where $b_0$ and $b_1$ are constants and 
$X_1=-\frac{\vec{P}_k^2\eta_1^2}{2a_1^4\left(6\delta-5\right)^2}$.
With $\alpha=6$ this leads to 
\begin{eqnarray}
\frac{B_k^2}{E_k^2}\simeq\mu_0
\left(\frac{\eta}{\eta_1}\right)^{-12(\delta-1)}
+\mu_1\left(\frac{\eta}{\eta_1}\right)^{-2}
+\mu_2\left(\frac{\eta}{\eta_1}\right)^{5-6\delta},
\end{eqnarray}
where $\mu_0$, $\mu_1$ and $\mu_2$ are constants 
which can be found from the expressions for $E_k^2$ and $B_k^2$.
Imposing the initial condition $E_k^2(\eta_2)\simeq B_k^2(\eta_2)$
and that all terms contribute equally at this time results in 
\begin{eqnarray}
\frac{B_k^2}{E_k^2}\simeq\frac{1}{3}\left(\frac{\eta_2}{\eta}
\right)^{12(\delta-1)}+\frac{1}{3}\left(\frac{\eta_2}{\eta}
\right)^2+\frac{1}{3}\left(\frac{\eta_2}{\eta}\right)^{6\delta-5}.
\end{eqnarray}
Thus in order to achieve, $B_k^2/E_k^2\leq 1$ the constant $b_0$ 
in equation (\ref{e341}) has to be set to zero.
With the remaining two terms contributing equally at $\eta=\eta_2$
and requiring $\frac{1}{2}<\delta<\frac{5}{6}$ leads to 
solutions which are consistent with the assumption $B_k^2/E_k^2\leq 1$.
Furthermore, in the expression for $B_k^2$ the dominant contribution comes
from the last term, thus the evolution of the magnetic field is given by
$\vec{B}_k^2\sim\left(\frac{\eta}{\eta_1}\right)^{\beta}$ where
$\beta=11-6\delta$ and $\frac{1}{2}<\delta<\frac{5}{6}$.
Moreover, the ratio of the energy density in the magnetic field
and the background radiation $r$ at the end of inflation can be 
calculated, resulting in 
\begin{eqnarray}
r(a_1)\simeq\left(9.2\times 10^{25}\right)^{-\beta}
\left(\frac{\lambda}{{\rm Mpc}}\right)^{-\beta}
\left(\frac{M}{M_P}\right)^{6-\frac{2\beta}{3}}
\left(\frac{T_{RH}}{M_P}\right)^{-2-\frac{\beta}{3}}.
\label{r_2}
\end{eqnarray}
In Figure 2 $\log r$ is shown. As can be seen the resulting magnetic
field strengths are far below the lower boundary of $r_0=10^{-37}$,
corresponding to a magnetic seed field of $B_s=10^{-20}$ G.
\begin{figure}[ht]
\centerline{\epsfxsize=3in\epsfbox{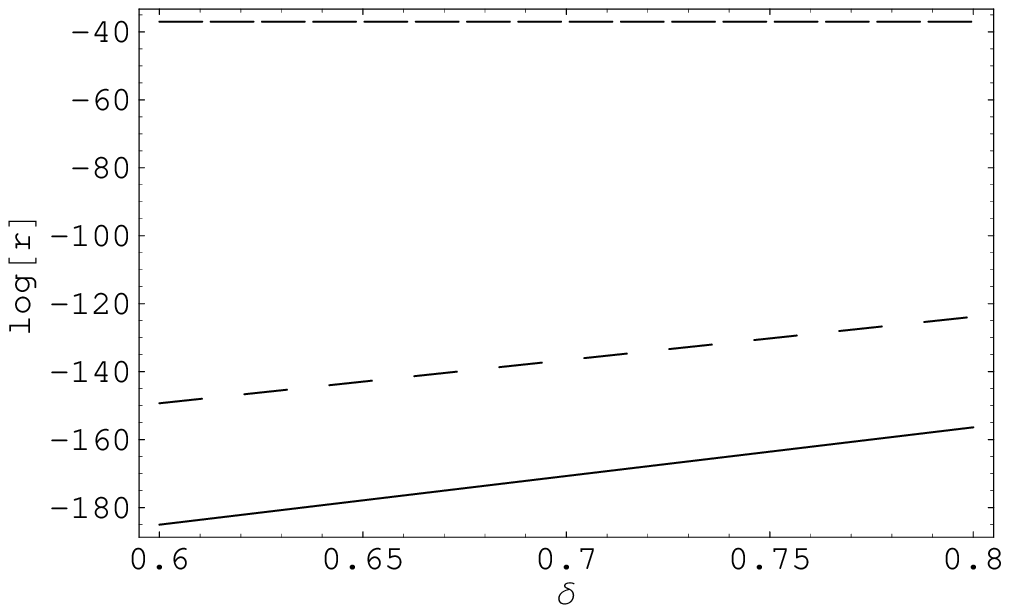}\hspace{1.0cm}
\epsfxsize=3in\epsfbox{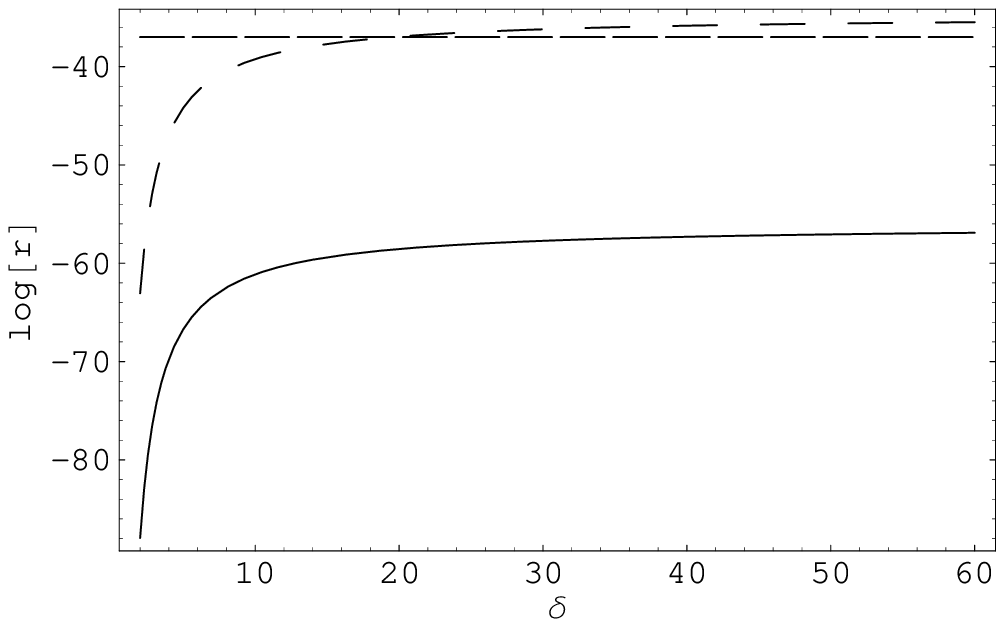}}
\caption{
For $\lambda=1$ Mpc $\log r$ (cf. equation (\ref{r_2}))  is shown as a 
function of $\delta$ for $T_{RH}=10^{17}$ GeV (black line) and $T_{RH}=10^{9}$ 
GeV (long-dashed line).
The dashed line corresponds to $r=10^{-37}$.
The left panel corresponds to the case $\vec{P}_k^2>0$.
The right panel corresponds to the case $\vec{P}_k^2=0$.}
\label{fig2}
\end{figure}

Finally, the solution for $\vec{P}^2_k\equiv 0$ will be discussed.
Thus using equation (\ref{e1}) and $X\simeq-\frac{1}{2}\vec{E}_k^2$
yields to
\begin{eqnarray}
\vec{E}_k^2\sim\left(\frac{\eta}{\eta_1}\right)^{\frac{4}{2\delta-1}}.
\end{eqnarray}
It is found that the solutions are consistent with the 
assumption $\Ec^{\,2}>\Bc^{\,2}$ for $1<\delta<\frac{3}{2}$
implying $\beta=4$ and for $\delta>\frac{3}{2}$ corresponding to 
$\beta=2\frac{2\delta+1}{2\delta-1}$. 
Thus using $\beta=4$ and $\lambda=1$ Mpc yields to 
$r(a_1)\simeq 10^{-104}$ for $M=10^{17}$ GeV and $T_{RH}=10^{17}$GeV.
Moreover $r(a_1)\simeq 10^{-77}$ is found for 
$M=10^{17}$ GeV and $T_{RH}=10^{9}$GeV.
These values are far below the lower bounds on the primordial 
magnetic field required to seed the galactic field.
The results for $\delta>\frac{3}{2}$ are shown 
in Figure 2.
As can be seen for $T_{RH}=10^9$ GeV 
magnetic fields satisfying $r>10^{-37}$ can be generated for
$\delta>19.5$.

\subsection{Discussion}

Solutions for the magnetic energy density of nonlinear electrodynamics with a lagrangian given by
$L=-\left(\frac{X^2}{\Lambda^8}\right)^{\frac{\delta-1}{2}}X$, where $\Lambda$
and $\delta$ are constant parameters have been found for different approximations. 
The solutions are determined by the system of equations (\ref{E1}) and (\ref{B1}).
These equations depend on two constants, $\vec{P}_k^2$ and $\vec{K}_k^2$.
In the case where $\vec{P}_k^2=0$, equation (\ref{E1}) is replaced by 
equation (\ref{e1}) which involves a new constant, $\vec{M_k}^2$, in the final 
equation (\ref{b4}). Furthermore, $\vec{M}_k^2$ and $\vec{K}_k^2$ lead to the definitions
of the two dimensionless constants $\alpha_1$ and $\alpha_2$ in equation (\ref{fode})
and $m^2\equiv\frac{\alpha_2}{\alpha_1}$. 

It is assumed that the electric and magnetic fields, respectively, have their origin
in quantum fluctuations during inflation. Therefore, it seems quite natural
to impose that initially, that is at the time when the perturbation leaves 
the horizon, the energy density in the electric and magnetic field are of 
the same order. During the later evolution these quantites of course can
be very different. In order to solve the equations, we have assumed 
three different types of evolution of the ratio of the energy densities
in the electric and magnetic field. This has led to estimates of 
the primordial magnetic field at the time of galaxy formation.

In the case $\Bc^{\,2}\simeq{\cal O}(\Ec^{\,2})$  for
$\vec{P}_k^2=0$ it was found that strong primordial magnetic fields
can be generated. 

Assuming that during inflation $\Bc^{\,2}\gg\Ec^{\,2}$
there is a range of the Pagels-Tomboulis parameter $\delta$ for which  in the case 
$\vec{P}_k^2>0$ primordial magnetic fields can be generated that are
strong enough to seed the galactic dynamo. In particular, 
for $\delta>1.9$ for $T_{RH}=10^{17}$ GeV and 
for $\delta>3.0$ for $T_{RH}=10^9$ GeV the ratio of the 
energy density of the magnetic field over
the energy density of the background radiation r is found to be
$r>10^{-37}$ corresponding to a primordial magnetic
field of at least $B_s=10^{-20}$G (cf. Figure 1). 
However, in the case $\vec{P}^2=0$ this solution is not consistent 
with the approximation $\Bc^{\,2}\gg\Ec^{\,2}$. Thus it cannot be used to
estimate the primordial magnetic field in this case. 
The former class of solutions do not inlude the case $\delta=\frac{1}{2}$.
In that case the solutions found for the electric and magnetic field
are consistent with the approximation for $\vec{P}_k^2>0$ and 
$\vec{P}_k^2=0$. Moreover, the resulting magnetic field is strong
enough to seed the galactic dynamo.

Finally, making the approximation $\Ec^{\,2}\gg\Bc^{\,2}$ 
yields in the case $\vec{P}_k^2>0$ to very weak magnetic fields. 
However, in the case $\vec{P}_k^2=0$, for $\delta>19.5$ and a 
reheat temperature $T_{RH}=10^9$ GeV  primordial magnetic fields
result which could successfully act as seed fields for 
the galactic dynamo (cf. Figure 2).

\section{Conclusions}

Observations of magnetic fields on large scales provide an 
intriguing problem. 
A possible class of mechanisms to create such fields 
is provided by inflationary models. Fluctuations in the 
electromagnetic field are amplified during 
inflation and provide a seed magnetic field at the time 
of structure formation which might be further 
amplified by a dynamo process. In general a sufficiently strong initial 
field strength can only be achieved if the conformal 
invariance of electrodynamics is broken. This has been realized,
for example, in models where the Maxwell lagrangian has been coupled 
to a scalar field, to curvature terms, etc. or by breaking Lorentz 
invariance or adding extra dimensions.

Here nonlinear electrodynamics has been considered. 
It has been assumed that whereas during the early universe 
electrodynamics is nonlinear it becomes linear 
at the end of inflation.
In particular the evolution of the magnetic energy density has been 
studied in a model of nonlinear electrodynamics which is described 
by a lagrangian of the form 
$L\sim -\left[\left(F_{\mu\nu}F^{\mu\nu}\right)^2/\Lambda^8\right]^{\frac{\delta-1}{2}}
F_{\mu\nu}F^{\mu\nu}$,
where $\Lambda$ and $\delta$ are  parameters. 
Originally the nonabelian version of this model had been 
proposed to describe low energy QCD \cite{pt}.
Here this model has been chosen as it is a strongly nonlinear theory of 
electrodynamics which allows to study in a semi-analytical way 
the possible creation and amplification of
a primordial magnetic field during de Sitter inflation. 
This is so since on the one hand the lagrangian only depends on 
one of the electromagnetic invariants, namely $X=\frac{1}{4}F_{\mu\nu}F^{\mu\nu}$,
which leads to a significant simplification of the equations.
On the other hand the power-law structure of the lagrangian make the equations
simpler. 

Approximate solutions have been found in three regimes of approximation
which describe the evolution of the ratio of the energy densities of
the electric and magnetic fields during inflation.
It is assumed that initially the energy density of the electric and
magnetic field are of the same order. Furthermore, these initial
fields are due to quantum fluctuations in the electromagnetic field
during inflation.
Whereas in the radiation dominated era, the energy density in the magnetic field
decreases as $a^{-4}$,  
the electric field strength rapidly decays in the highly conducting plasma
(see, e.g., \cite{tw,d93}).
Solutions in closed form have been found and the resulting primordial 
magnetic field estimated.
It has been shown that depending on the regime of approximation and 
the value of the Pagels-Tomboulis parameter $\delta$
primordial magnetic fields can be generated that are strong enough
to seed a galactic dynamo.
Thus we have provided an example of a theory of nonlinear electrodynamics 
where the nonlinearities act in a way as to amplify sufficiently an initial magnetic
field.

The energy-momentum tensor of the electromagnetic field
can be cast in the form of an imperfect fluid. This has been found explicitly for the 
particular model of nonlinear electrodynamics under consideration here.
Moreover, this allows to find the expression for the energy density $\rho$ of 
the fluid. Requiring that $\rho$ should be positive provides the bound
$\delta\geq\frac{1}{2}$.

In  \cite{gfc} the possible creation and amplification of magnetic fields was studied in an inflationary model coupled to a pseudo Goldstone boson (see also \cite{tw}). In this case
the lagrangian has the form $L\sim\frac{1}{2}\partial_{\mu}\theta\partial^{\mu}\theta
-X+g_a\theta Y$, where $\theta$ is the axion field. This provides an example of 
a more general lagrangian having also an explicit dependence on 
$Y=\frac{1}{4}F_{\mu\nu}\;^{*}F^{\mu\nu}$. 
However, as it turns out the resulting primordial magnetic field is not strong enough 
in order to seed, for example, a galactic dynamo.
In \cite{as} the model of \cite{gfc} was generalized to N axions. In this case it was found that 
at least the weaker bound of $r>10^{-57}$ can be satisfied.
Here, in this work the creation of primordial magnetic fields in a particular model
of nonlinear electrodynamics has been studied. It might be interesting to 
generalize this to lagrangians depending on both electromagnetic invariants
$X$ and $Y$.

\section{Acknowledgements}

I would like to thank M. A. V\'azquez-Mozo for enlightening discussions.
I am grateful to the CERN theory division for hospitality where 
part of this work was done.
This work has been supported by the ``Ram\'on y Cajal'' programme of 
the MEC (Spain). Partial support by Spanish Science Ministry grants
FPA2005-04823 and FIS2006-05319 is gratefully acknowledged.

\section{Appendix A}
\setcounter{equation}{0}

In this section it is checked that the approximate exact solution (\ref{b5}) 
is a good approximation to the solution of the full differential 
equation (\ref{fode}).
The solution (\ref{b5}) satisfies equation (\ref{apode}). Writing the 
full differential equation (\ref{fode})
as
\begin{eqnarray}
y\ddot{y}=\delta\dot{y}^2+\frac{m^2}{1-\delta}y^2+I,
\end{eqnarray}
where for the approximate solution $y=C_2\cosh(z)^{\frac{1}{1-\delta}}$
with $z\equiv m(x+(\delta-1)C_1)$
the additional term $I$ is given by
\begin{eqnarray}
I&\equiv&\frac{C_2^{2\delta+1}}{\alpha_1(\delta-1)}\cosh^{\frac{2\delta+1}{1-\delta}}(z)\left(\frac{x}{x_1}\right)^{-4}\left[
(2\delta-1)\frac{m^2}{(1-\delta)^2}\tanh^2(z)
\right.
\nonumber\\
&-&
\left.\frac{4(\delta+1)}{x_1}
\left(\frac{x}{x_1}\right)^{-1}\frac{m}{1-\delta}\tanh(z)
+\frac{20}{x_1^2}\left(\frac{x}{x_1}\right)^{-2}+\frac{m^2}{1-\delta}
\right].
\end{eqnarray}
At $x_2$ when the comoving length scale $\lambda$ leaves the horizon
$z=0$ by construction. Thus $I$ is 
proportional to $\left(\frac{x_2}{x_1}\right)^{-4}\ll 1$.
At the end of inflation, $x=x_1$, using the bound on $-mx_1$ which 
in general implies $-mx_1\ll 1$, $I(x_1)$ is given approximately by
\begin{eqnarray}
I(x_1)\sim\frac{20C_2^{2\delta+1}}{(\delta-1)\alpha_1x_1^2}
\cosh^{-\frac{2\delta+1}{\delta-1}}(z_1),
\end{eqnarray} 
where the last factor is much less than 1 since it is assumed that 
$\delta>1$ and, moreover, $z_1\sim-mx_1 e^{N(\lambda)}\gg 1$.
Thus choosing $C_2$ appropriately, $|I(x_1)|\ll 1$.

Finally, it can also be checked using the bounds on $-mx_1$ that 
the square of the  magnetic field strength $\vec{B}^2_k$ is well approximated by 
equation (\ref{B-app}).

\section{Appendix B}
\setcounter{equation}{0}

Expressions for $\frac{L_X'}{L_X}$ and $\frac{L_X''}{L_X}$ for the solution 
(\ref{B0.5}).
\begin{eqnarray}
\frac{L_X'}{L_X}&=&-\frac{2}{\eta_1}\left(\frac{\eta}{\eta_1}\right)^{-1}-3\frac{\mu}{\eta_1}
\left(\frac{\eta}{\eta_1}\right)^2\tanh\left[\mu\left(\frac{\eta}{\eta_1}\right)^3-c_1\right]\\
\frac{L_X''}{L_X}&=&\frac{6}{\eta_1^2}\left(\frac{\eta}{\eta_1}\right)^{-2}
-6\frac{\mu}{\eta_1^2}
\left(\frac{\eta}{\eta_1}\right)\tanh\left[\mu\left(\frac{\eta}{\eta_1}\right)^3-c_1\right]
\nonumber\\
&&
+9\frac{\mu^2}{\eta_1^2}\left(\frac{\eta}{\eta_1}\right)^4-18\frac{\mu^2}{\eta_1^2}
\left(\frac{\eta}{\eta_1}\right)^4\cosh^{-2}\left[\mu\left(\frac{\eta}{\eta_1}\right)^3-c_1\right],
\end{eqnarray}
where $\mu\equiv-\left(\frac{\alpha_2}{18}\right)^{\frac{1}{2}}x_1$.

\end{document}